\documentclass[12pt]{article}
\usepackage{graphicx}

\begin{document}

\centerline{\large \bf MODELS OF REALISTIC REGGEONS}
\centerline{\large \bf FOR t=0 HIGH ENERGY SCATTERING}

\vskip 0.7cm

\centerline{J. Kontros$^{\ast}$, K. Kontros$^{\dagger}$, A.
Lengyel$^{\ddagger}$}

\vskip .3cm

\centerline{\sl Institute of Electron Physics, Universitetska 21,}
\centerline{\sl 88000 Uzhgorod, Ukraine}

\vskip .5cm

\begin{abstract}
Several Regge-phenomenological models of forward scattering at
high energies are considered. It was found that one of them
corresponds to the form of finite series in $\log{s}$ with
additional terms that represent the exchanges of non-degenerate
Regge-trajectories. These trajectories are unique for both the
scattering and resonance regions and satisfy the most realistic
ideas on them.
\end{abstract}

\section{Introduction}

Since the middle of the last century Pomeron plays a fundamental
role in describing the scattering at high energies. In spite of
this, our understanding of the Pomeron nature is far from
completeness and requires further specifications. At the same time
the secondary Reggeons were undeservedly moved to the role
reflected in their name.

In this paper, we will consider the Regge-phenomenological models
of forward scattering at high energies, where both Pomeron and
secondary Reggeons (further on Reggeons) satisfy the most
realistic ideas on them.

From the Regge-theory point of view the basic characteristic of
Pomeron and Reggeons is the intercept \cite{BGJPP}. This concept
has the most effective and economic form in the
Donnachie-Landshoff (DL) model \cite{DL}, which we considered as a
standard reference for the models of total, elastic and
diffractive cross sections. For the $pp$ and $\overline{p}p$
scattering the total cross-section in this model is:

\begin{equation}
\label{eq1} \sigma _{tot}(s) =Xs^{\varepsilon}+ Y_{\pm}s^{-\eta},
\end{equation}
where $Y_{\pm}$ corresponds to $pp$ and $\overline{p}p$ scattering
with exchange degenerate Reggeon representing both $C=\pm
1(\rho,\omega,a_{2},f)$ exchanges besides the Pomeron with
intercepts given by $\alpha_{P}=1+\varepsilon$ and
$\alpha_{R}=1-\eta$. The DL model fares reasonably well when
fitting to the $pp$ and $\overline{p}p$ total cross sections and
demands a generalization when fitting both the total cross
sections and the real parts of the scattering amplitudes
\cite{CMG,CKK}. Then the $pp$ and $\overline{p}p$ scattering of
the total cross sections in this model will be:

\begin{equation}
\label{eq2} \sigma_{tot}(s)=Xs^{\varepsilon}+Y_{+}s^{-\eta_{+}}
\mp Y_{-}s^{-\eta_{-}},
\end{equation}
where the last two terms represent the exchanges of non-degenerate
$C=+1(a_{2},f)$ and $C=- 1(\rho,\omega)$ trajectories with
intercepts $\alpha_{\pm}=1-\eta_{\pm}$, respectively. The sign of
$Y_{-}$ term flips when fitting $pp$ data are compared to the
$\overline{p}p$ data.

\vskip .5cm

\hrule

\vskip .2cm

\noindent

\vfill

$ \begin{array}{ll}^{\ast,\dagger}\mbox{{\it e-mail
address:}} &
 \mbox{jeno@kontr.uzhgorod.ua}
\end{array}
$

$ \begin{array}{ll}^{\ddagger}\mbox{{\it e-mail address:}} &
 \mbox{sasha@len.uzhgorod.ua}
\end{array}
$

It is well known that models (\ref{eq1}) and (\ref{eq2}) violate
the unitarity. As it was pointed out in \cite{CMG}, the unitarity
violation occurs at the energies only slightly above the Tevatron
energy of $1.8$ $TeV$, and therefore it is a problem of the
present and not of the future. Not less successful models are
those based on the more complex analytical properties of the
scattering amplitude, not violating however the unitarity
conditions \cite{DGLM,JRC}:

\begin{equation}
\label{eq3} \sigma _{tot}(s)=A + B\log{s} + Y_{+} s^{-\eta_{+}}
\mp Y_{-} s^{-\eta_{-}},
\end{equation}

\begin{equation}
\label{eq4} \sigma_{tot}(s)=A + B\log{^{2}s} + Y_{+} s^{-\eta_{+}}
\mp Y_{-} s^{-\eta_{-}}.
\end{equation}

An alternative way of effective account of the complex structure
of the singularities is to try to mimic them by a two-component
Pomeron built from two Regge singularities \cite{DGLM,BLV,GN}.
Note that the perturbative Pomeron has also a complex form.
Recently detailed calculations in QCD indicated an existence of
the two-component Pomeron \cite{BLV}.

The two-component Pomeron model for the total cross sections has
the following form:

\begin{equation}
\label{eq5} \sigma _{tot}(s)=Z+Xs^{\varepsilon}+Y_{+}
s^{-\eta_{+}} \mp Y_{-} s^{-\eta_{-}},
\end{equation}
where the second component corresponds to the intercept larger
than 1 ($\varepsilon>0$) and the first component corresponds to an
intercept exactly localized at 1.

A new development of this model supposes that the X-component is
fully universal, i.e. its coupling is the same in all
hadron-hadron reactions \cite{GN,DN}, while the first one is a
non-universal Pomeron.

Another generalization of the multicomponent Pomeron is based on
the assumption that QCD Pomeron corresponds to the infinite sum of
gluon ladders with Reggeized gluons on the vertical lines
\cite{BL}. Essentially at finite energies only a finite number of
diagrams contributes, giving rise to a finite series in $\log{s}$
\cite{FJKLPP} like:

\begin{equation}
\label{eq6} \sigma _{tot}(s)=A+B\log{s}+C\log{^{2}s}+Y_{+}s^{-\eta
_{+}} \mp Y_{-} s^{-\eta_{-}}.
\end{equation}

Because of the fact that most of mentioned models have been tested
with different data sets, it is very difficult to compare between
them. Fortunately there exists a complete data set of total cross
sections and ratios of the real part of scattering amplitude to
the imaginary part for the $pp$, $\overline{p}p$, $\pi^{\pm}p$,
$K^{\pm}p$ scattering and for total cross sections in the case of
the $\gamma p$, $\gamma \gamma$ and $\Sigma p$ scattering
\cite{DATA}. It was found in \cite{JRC} with the data set
\cite{DATA} that the data cannot discriminate between a
simple-pole fit (\ref{eq2}) and asymptotic $\log{s}$ (\ref{eq3})
and $\log{^{2}s}$ (\ref{eq4}) fits. The models examined in
\cite{JRC} satisfy the condition $\chi^{2}/dof\approx1$ with 16
fitted parameters at $\sqrt{s}_{min} \ge 9GeV$. According to the
program initiated in \cite{JRC} by scrutinizing two earlier
unexplored models for the Pomeron (\ref{eq5}) and (\ref{eq6}) with
the weak degeneration as in (\ref{eq1}) for $\sqrt{s}_{min} \ge
10GeV$ it was found in \cite{KKL} that the condition
$\chi^{2}/dof\sim1$ satisfies with the same number of fitted
parameters.

The analysis of all mentioned models shows that the values of
fitted intercepts of Reggeons sufficiently differ (see Fig. 1).
Moreover, the fitted values of intercepts are far from the
available data for the corresponding mesonic resonances drawn on
the Chew-Frautschi plot. Therefore a fairly conclusive analysis
can be performed using, on the one hand, the data in the resonance
region, and, on the other hand, those for forward scattering
\cite{DGMP}.

\begin{figure}[ht]
\begin{center}
\includegraphics*[scale=0.4]{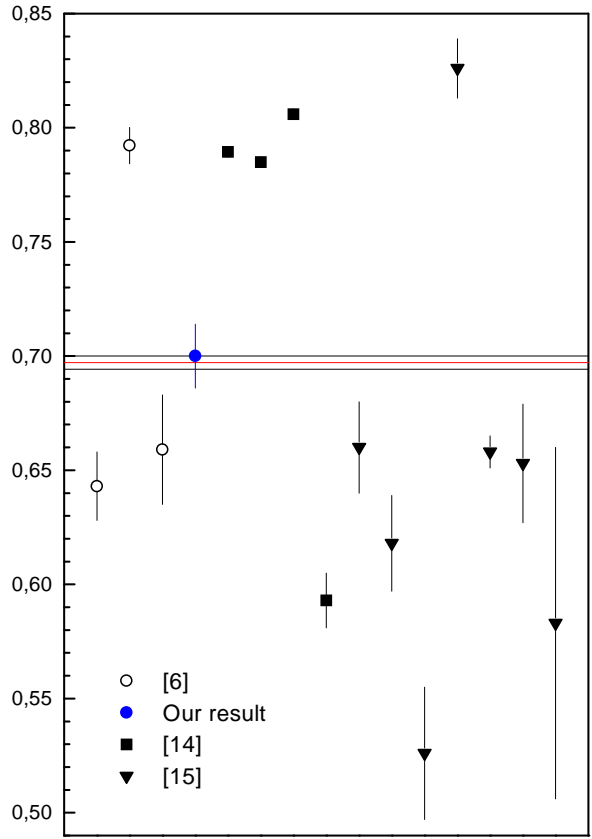}
\caption[]{\small Comparison of the $f$-Reggeon intercept values
obtained from different models. The red line and the two black
lines correspond to the $f$-Reggeon intercept value and its error
corridor calculated from the Chew-Frautschi plot \cite{DGMP}.}
\end{center}
\end{figure}

Below we suggest a Regge-phenomenological model (\ref{eq6}) with
both Pomeron and Reggeons satisfying the most realistic notions
about them. It will be demonstrated that Pomeron as a finite sum
of $\log{s}$ terms (\ref{eq6}) is quite enough, the coefficients
of this series decrease with respect to each other as $\sim 1/5$
and $\sim 1/15$.

Nevertheless, at simultaneous fit of Pomeron and Reggeons
parameters the values of the Reggeons intercepts coincide with the
values obtained from Chew-Frautschi plot. In addition note that
the $\chi^{2}/dof\sim1$ criterion is reached at $\sqrt{s}_{min}=8
GeV$ energy.

Furthermore we study how compatible are the above mentioned models
with the realistic interpretation of the unique Regge trajectory
for both the scattering and resonance regions as well. Finally on
the basis of this criterion we define the realistic intercept
values of the supercritical Pomeron.

\section{Finite $\log{s}$ series for the Pomeron}

Let us write the real and imaginary parts of the forward elastic scattering
amplitude as

\begin{equation}
\label{eq7} \frac{1}{s}ImA_{h_{1} h_{2}} (s)=ImP(s) + ImR(s),
\end{equation}

\begin{equation}
\label{eq8} \frac{1}{s}ReA_{h_{1} h_{2}} (s)=ReP(s)+ReR(s),
\end{equation}
where

\begin{equation}
\label{eq9} ImP(s)=\lambda_{h_{1} h_{2}}  \left[A+B\log{\left(
{\frac{s}{s_{0}}}\right)}+C\log{^{2}\left({\frac{s}{s_{0}}}
\right)} \right],
\end{equation}

\begin{equation}
\label{eq10} ReP(s)=\frac{\pi}{2}\lambda_{h_{1} h_{2}} B + \pi
\lambda _{h_{1} h_{2}} C\log{\left( {\frac{s}{s_{0}}} \right)},
\end{equation}
come from the Pomeron contribution and

\begin{equation}
\label{eq11} ImR(s)=Y_{1}^{h_{1} h_{2}} s^{-\eta_{1}} \mp
Y_{2}^{h_{1} h_{2}} s^{-\eta_{2}},
\end{equation}

\begin{equation}
\label{eq12} ReR(s)=\left[ {Y_{1}^{h_{1} h_{2}}
s^{-\eta_{1}}\cot{\left( {\frac{1-\eta_{1}}{2}\pi}  \right)} \mp
Y_{2}^{h_{1} h_{2}} s^{-\eta _{2}} \tan{\left( {\frac{1-
\eta_{2}}{2}\pi} \right)}} \right],
\end{equation}
correspond to the contribution of two non-degenerate Regge
trajectories.

We have fitted the above model to the data on the total cross
sections and the ratio of the real to imaginary part of $pp$,
$\overline{p}p$, $\pi^{\pm}p$, $K^{\pm}p$, $\gamma p$ and $\gamma
\gamma$ as well as the $\Sigma p$-scattering, starting from
$\sqrt{s}_{min}=3GeV$ up to the highest available energies.
Similarly to ref. \cite{JRC}, we have studied the stability of our
fit by varying the lower limit of the fit $\sqrt{s}_{min}$ between
3 and 13 $GeV$. The resulting fit (value of $\chi^{2}/dof$) as
well as the dependence of one of the fitted parameters on the
lower bound $\sqrt{s}_{min}$ are presented in Fig. 2 and Fig. 3.

\begin{figure}[ht]
\begin{center}
\includegraphics*[scale=0.4]{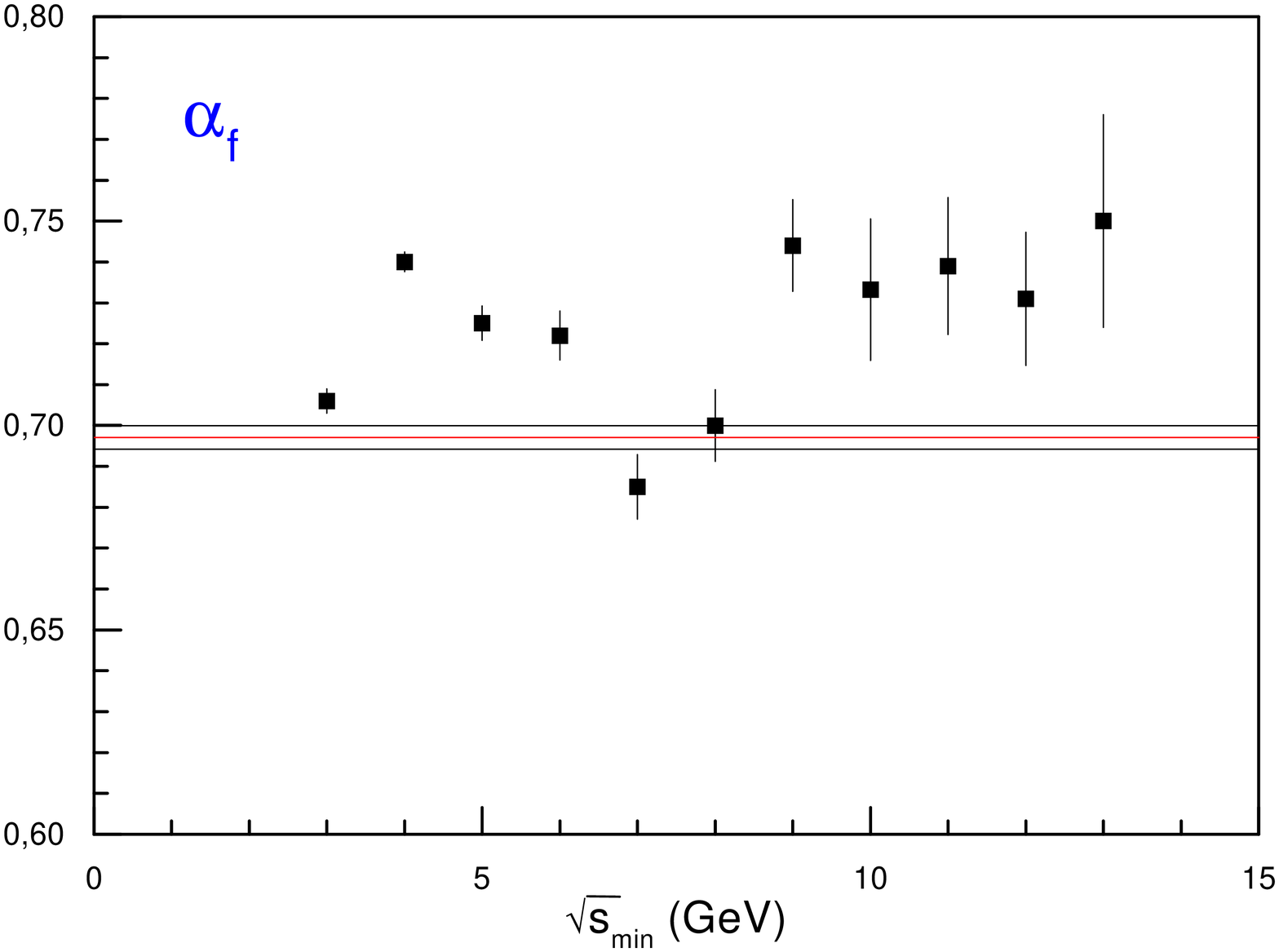}
\caption[]{\small Values of the $f$-Reggeon intercept as a
function of the minimum energy}{\small considered in the fit. The
lines are the same as in Fig. 1.}
\end{center}
\end{figure}

The result of our fits is good as those of ref. \cite{JRC}. The
limiting value $\chi^{2}/dof \approx 1$ is reached at
$\sqrt{s}_{min}=8GeV$.

An important result of our fit is that the coefficients of the
finite series of Pomeron (\ref{eq9}) (see Table 1) decrease in
their absolute values (as $\sim 1/5$ and $\sim 1/15$) providing a
fast convergence of the series and ensuring the applicability of
this approximation at still much higher energies. The physical
motivation of such a finite series representation of the Pomeron
was discussed earlier - both in the context of Regge multipoles
and QCD \cite{BL,FJKLPP}.

\bigskip

{\scriptsize
\begin{tabular}{|ccccccc|}
\hline \multicolumn{1}{|c|}{$A$ (mb)} & \multicolumn{1}{c|}{$B$
(mb)} & \multicolumn{1}{c|}{$C$ (mb)} &
\multicolumn{1}{c|}{$\eta_{1}$} & \multicolumn{1}{c|}{$\eta_{2}$}
& \multicolumn{1}{c|}{$\chi ^2$/d.o.f.} & statistics \\ \hline
\multicolumn{1}{|c|}{$10.5\pm 0.7$} &
\multicolumn{1}{c|}{$2.201\pm 0.076$} &
\multicolumn{1}{c|}{$0.138\pm 0.011$} &
\multicolumn{1}{c|}{$0.300\pm 0.015$} &
\multicolumn{1}{c|}{$0.563\pm 0.013$} & \multicolumn{1}{c|}{$1.0$}
& $453$
\\ \hline \multicolumn{1}{|c|}{} & \multicolumn{1}{c|}{$pp$} &
\multicolumn{1}{c|}{$ \pi p$} & \multicolumn{1}{c|}{$Kp$} &
\multicolumn{1}{c|}{$\Sigma p$} & \multicolumn{1}{c|}{$\gamma
p\times 10^{-3}$} & $\gamma \gamma \times 10^{-4}$ \\ \hline
\multicolumn{1}{|c|}{$\lambda $} & \multicolumn{1}{c|}{$1$} &
\multicolumn{1}{c|}{$0.6552\pm 0.0037$} &
\multicolumn{1}{c|}{$0.5912\pm 0.0055$} &
\multicolumn{1}{c|}{$0.99\pm 0.03$} &
\multicolumn{1}{c|}{$3.000\pm 0.046$} & $0.091\pm 0.006$ \\
\multicolumn{1}{|c|}{$Y_1$ (mb)} & \multicolumn{1}{c|}{$69.43\pm
1.73$} & \multicolumn{1}{c|}{$32.59\pm 0.53$} &
\multicolumn{1}{c|}{$21.1\pm 0.4$} & \multicolumn{1}{c|}{$37.4\pm
4.1$} & \multicolumn{1}{c|}{$144\pm 5$} & $4.0\pm 1.2$ \\
\multicolumn{1}{|c|}{$Y_2$ (mb)} & \multicolumn{1}{c|}{$36.6\pm
2.4$} & \multicolumn{1}{c|}{$7.35\pm 0.54$} &
\multicolumn{1}{c|}{$14.7\pm 1.0$} & \multicolumn{1}{c|}{} &
\multicolumn{1}{c|}{} &  \\ \hline
\end{tabular}
}

\medskip

Table 1: Values of the fitted parameters in the finite series
Pomeron model (\ref{eq7})-(\ref{eq12}),

with a cut-off $\sqrt{s}_{min}=8 GeV$ and $s_{0}=1 GeV^{2}$ fixed.

\bigskip

However, this is not a single advantage of our model. Note that in
the process of the fit the obtained intercepts of Reggeons well
coincide with those calculated from the Chew-Frautschi plot
\cite{DGMP}. As is seen in Fig. 1, our model belongs to the very
few models that give realistic values of the intercept for the
$f$-Reggeon. Regarding its quantum numbers the $f$-Reggeon is
inseparable from Pomeron and thus the definition of its parameters
is being model dependent.

\begin{figure}[ht]
\begin{center}
\includegraphics*[scale=0.4]{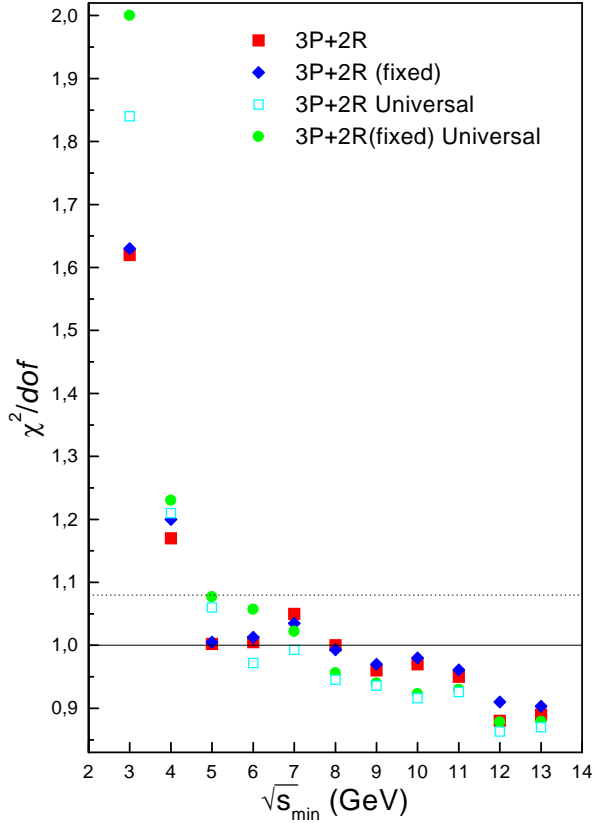}
\caption[]{\small $\chi^{2}/dof$ as the function of the minimum
energy considered in the fit: 3P+2R - model (\ref{eq6}) with both
Pomeron and Reggeons parameters fitted; 3P+2R (fixed) - model
(\ref{eq6}) with Reggeons intercept values (\ref{eq13}); 3P+2R
Universal - model (\ref{eq14}) with both Pomeron and Reggeons
parameters fitted; 3P+2R (fixed) Universal - model (\ref{eq14})
with Reggeons intercept values (\ref{eq13}).}
\end{center}
\end{figure}

The intercept value of $\omega$-Reggeon is in a quite good
agreement with the values taken from the Chew-Frautschi plot
\cite{DGMP}, despite of the simplicity of the calculations based
on the degeneration of $f/a$ and $\omega/\rho$. Taking into
consideration these properties of the model (\ref{eq6}) concerning
the Reggeons parameters, one can estimate the realistic parameters
of the Pomeron in a form of model (\ref{eq6}), if the parameters
of Reggeons are fixed, in accordance with the values obtained from
the fit of Chew-Frautschi plot \cite{DGMP}, still requiring the
stability conditions to hold $\chi^{2}/dof\sim 1$. To do this we
have chosen the following Reggeons intercept values \cite{DGMP}:

\begin{equation}
\label{eq13} \alpha_{f}=0.6971, \quad \alpha_{\omega}=0.4359,
\quad \alpha_{\rho}=0.4783,
\end{equation}
and here two Reggeons still contribute to the scattering
amplitude:

\begin{center}
$f$ and $\omega$ for $pp$, $\overline{p}p$, $K^{\pm} p$, $\Sigma
p-$scattering
\end{center}

\begin{center}
$f$ and $\rho$ for $\pi^{\pm} p-$scattering
\end{center}

\begin{center}
$f$ for $\gamma p$, $\gamma \gamma-$scattering
\end{center}

Our motivation is as follows. As shown in \cite{KKL}, the
contribution of $a_{2}$-meson $\approx 0$ and the contribution of
$\rho$-meson to the total cross sections of $pp$, $\overline{p}p$
and $K^{\pm} p-$scattering is comparable with the error of the
$\omega$-Reggeon contribution. Moreover, we have not included to
the fit the scattering data for neutral particles, for which the
contribution of these Reggeons is very important. The result of
the fit is shown in Fig. 3. As is seen from these data, the model
gives a slightly better result than the previous ones with free
parameter's fit, due to the fact that the number of parameters is
reduced by 2 and, that is essential, our suggestions about the
realistic model (\ref{eq6}) are valid.

\begin{figure}[ht]
\begin{center}
\includegraphics*[scale=0.4]{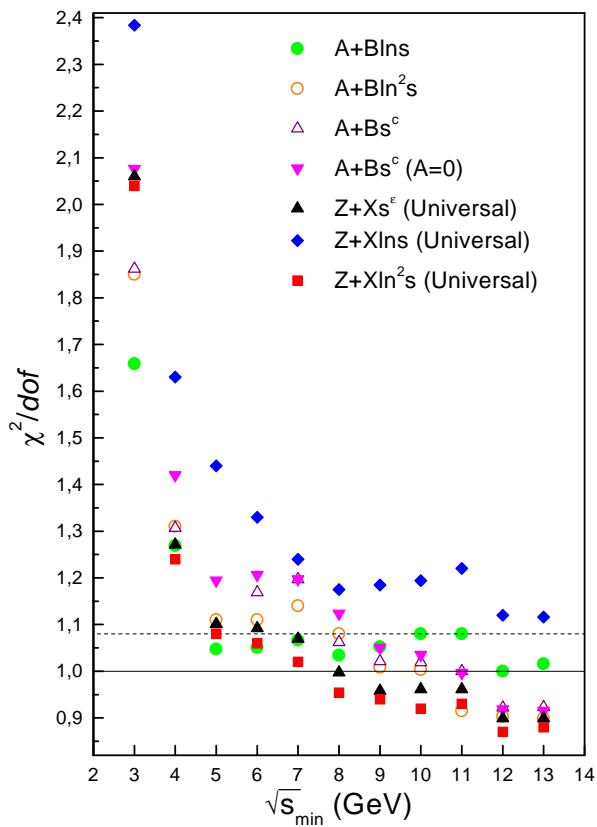}
\caption[]{\small $\chi^{2}/dof$ as the function of the minimum
energy considered in the fit: The first four models correspond to
the (\ref{eq2})-(\ref{eq5}) models with fixed Reggeons parameters
from (\ref{eq13}); The last three models correspond to the model
(\ref{eq14}) with the assumption of the G-universality and fixed
Reggeons intercept values (\ref{eq13}).}
\end{center}
\end{figure}

Recently it has been shown in \cite{BN} that the inclusion of
G-universality improves sufficiently some models describing the
scattering at t=0. We have included into our model the assumption
about the G-universality that now has a form:

\begin{equation}\label{eq14}
  \begin{array}{ccc}
    \sigma_{\overline{p}p}^{pp}=A_{pp}+P(s)+R_{\overline{p}p}^{pp}(s), \\
    \sigma_{K^{\pm}p}=A_{Kp}+P(s)+R_{K^{\pm}p}(s), \\
    \sigma_{\pi^{\pm}p}=A_{\pi p}+P(s)+R_{\pi^{\pm}p}(s), \\
    \sigma_{\gamma p}=\delta A_{pp}+\delta P(s)+f_{\gamma p}(s), \\
    \sigma_{\gamma \gamma}=\delta^{2} A_{pp}+\delta^{2} P(s)+f_{\gamma \gamma}(s), \\
    P(s)=B\log{s}+C\log{^2{s}}, \\
    R_{\overline{p}p}^{pp}(s)=f_{pp}\mp\omega_{pp},\\
    R_{K^{\pm}p}(s)=f_{Kp}\mp\omega_{Kp}, \\
    R_{\pi^{\pm}p}(s)=f_{\pi p}\mp\rho_{\pi p}.
\end{array}
\end{equation}

As is seen from the comparison of the fitting results, the latter
model demonstrates a stable improvement at all energies.
Unfortunately it is not valid for the $f$-Reggeon intercepts.

As one may expect, the version of our model with the assumption of
the G-universality and fixed Reggeons values gives the absolutely
best description of the whole experimental data set used here (see
Fig. 3).

If one can accept the reasonable criterion $\chi^{2}/dof \leq
1.08$ then as is seen from Fig. 3 all the four versions of our
model satisfy this criterion yet at the lower limit of the fit
$\sqrt{s}_{min}=5 GeV$.

\section{Scattering models at t=0 with realistic values
of Reggeons intercept}

It is seen from Fig. 1 that the $f$-Reggeon intercept values are
distinctly model-dependent and are far from the agreement with
those obtained from the Chew-Frautschi plot. Here we shall check
the criteria of applicability of the models under discussion if
one uses the fixed Reggeons intercept values from the
Chew-Frautschi plot, similarly to the previous section. The
results of such check are shown in Fig. 4. It is interesting to
note that most of the models obey the reasonable requirement $\chi
^{2}/dof \leq 1.08$ for $\sqrt{s}_{min}=8 GeV$, while for the four
versions of our model (\ref{eq6}) this condition is valid at
$\sqrt{s}_{min}=5 GeV$ (see previous section). The severe
requirement $\chi ^{2}/dof \leq 1$ is satisfied by the Pomeron
models for the boundary energy value of $\sqrt{s}_{min} \geq 10
GeV$.

\begin{center}
$P(s)=As^{\varepsilon}$ \quad (1) \quad \quad \quad \quad \quad
\quad $P(s)=A+B\log{^{2}s}$ \quad (3) \\ $P(s)=A+B\log{s}$ \quad
(2) \quad \quad \quad \quad \quad $P(s)=A+Bs^{\varepsilon}$ \quad
(4)
\end{center}

Among the model versions with the G-universality all the above
models except for (\ref{eq2}) obey this criterion.

\begin{figure}[ht]
\begin{center}
\includegraphics*[scale=0.4]{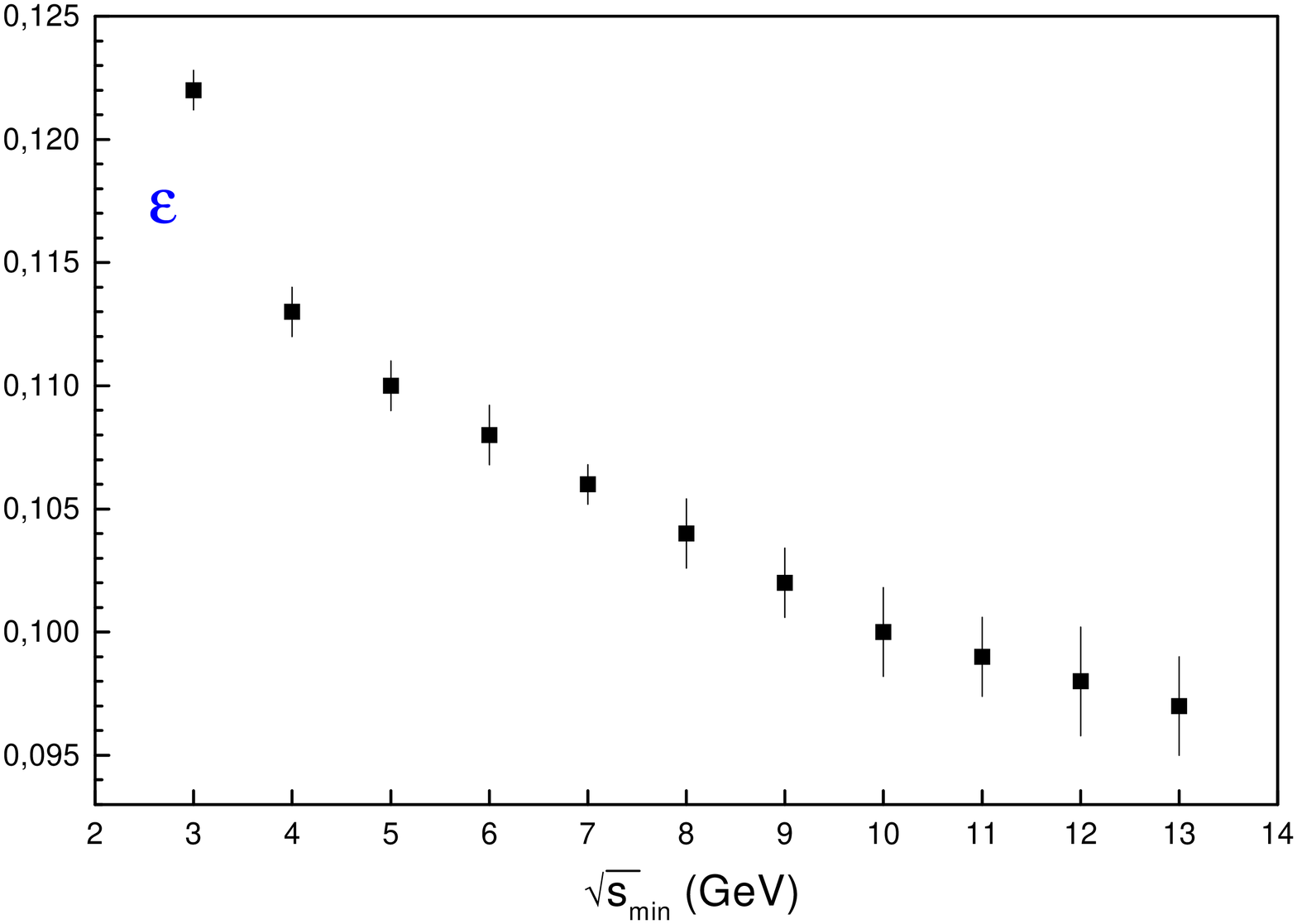}
\caption[]{\small $\varepsilon$ parameter of the supercritical
Pomeron}{\small as a function of the minimum energy considered in
the fit.}
\end{center}
\end{figure}

\section{Realistic intercept values for the Pomeron}

Since Donnachie-Landshoff determined the Pomeron intercept value
equal to $1.08$, debates are still around it. One of the reason is
in the unusual behavior of the Pomeron intercept in models of DIS,
where at $Q^{2}\rightarrow0$ the intercept value of the Pomeron is
close to the value obtained from the scattering region, and at
$Q^{2}\rightarrow\infty$ it drastically increases up to $1.3-1.4$.
The situation became dramatic, because the intercept value of the
BFKL Pomeron is also large. Scrupulous estimations for the Pomeron
intercept values were carried out in \cite{CMG,CKK}. It was found
that within the framework of the model (\ref{eq2}) the Pomeron
intercept equals to $1.104\pm0.002$ and $1.096^{+0.012}_{-0.009}$.
Herein we refitted the Pomeron intercept values in the spirit of
\cite{JRC} at fixed values of Reggeons (\ref{eq13}) for boundary
energies $3-13 GeV$ utilizing the same data set \cite{DATA}. At
$\chi ^{2}/dof=1.05$ ($\sqrt{s}_{min}=9 GeV$) the intercept value
is equal to $1.1020\pm0.0014$, while at $\chi ^{2}/dof=1$
($\sqrt{s}_{min}=11 GeV$) it is $1.0989\pm0.0016$. Calculations
are shown in Fig. 5. As is seen our result well coincides with the
previous calculations \cite{CMG,CKK}.

\section{Conclusions}

Summarizing the results and checking the Regge-phenomenological
models it is possible to formulate reasonably criteria for its
application:

\begin{enumerate}
  \item The check of models always should be done by using the same
data set (for example \cite{DATA}).
  \item The criterion of $\chi ^{2}/dof\leq1$ should be fulfilled.
  \item The compatibility of parameters with the secondary Reggeons from the resonance region.
\end{enumerate}

As the result of our analysis we have concluded that the most
realistic model includes the Pomeron in the form of finite series
in $\log{s}$, satisfying the Froissart bound and two Reggeons for
the best description of the forward scattering. In this model the
criterion $\chi ^{2}/dof\leq1$ was reached yet at the energy of
$\sqrt{s}_{min}=8 GeV$. Note that only this model gives the
$f$-Reggeon intercept value that agrees with those extracted from
the resonance region. The two-component Pomeron model (\ref{eq5})
and the Froissaron model (\ref{eq4}) with its G-universality
versions satisfy the above criteria beginning from
$\sqrt{s}_{min}=9 GeV$, the DL model (\ref{eq2}) beginning from
$\sqrt{s}_{min}=11 GeV$ and the dipole Pomeron model just from
$\sqrt{s}_{min}=12 GeV$. If one will change the criterion 2. to
the reasonable criterion $\chi ^{2}/dof \approx 1.08$ then
practically all the models are fulfilled.

\bigskip

The authors are grateful to L. Jenkovszky and E. Martynov for
useful discussions and support.


\begin{thebibliography}{99}
\bibitem{BGJPP} M. Bertini, M. Giffon, L. Jenkovszky, F. Paccanoni, E.
Predazzi, \textit{Riv. Nuovo Cim.} \textbf{19} (1996) 1.

\bibitem{DL} A. Donnachie, P.V. Landshoff, \textit{Phys. Lett.} \textbf{B296}
(1992) 227.

\bibitem{CMG} R.J.M. Covolan, J. Montanha, K. Goulianos, \textit{Phys. Lett.}
\textbf{B389} (1996) 176.

\bibitem{CKK} J.R. Cudell, K. Kang, S.K. Kim, \textit{Phys. Lett.}
\textbf{B395} (1997) 311.

\bibitem{DGLM} P. Desgrolard, M. Giffon, A. Lengyel, E. Martynov, \textit{Nuovo Cim.}
\textbf{107A} (1994) 637.

\bibitem{JRC} J.R. Cudell, V. Ezhela, K. Kang, S. Lugovsky, N.
Tkachenko, \textit{Phys Rev.} \textbf{D61} (2000) 034019.

\bibitem{BLV} J. Bartels, L.N. Lipatov, G.P. Vacca, \textit{Phys. Lett.} \textbf{B477}
(2000) 178.

\bibitem{GN} P. Gauron, B. Nicolescu, \textit{Phys. Lett.} \textbf{B486} (2000) 71.

\bibitem{DN} L.G. Dakhno, V.A. Nikonov, \textit{Eur. Phys. J.} \textbf{A5} (1999) 209.

\bibitem{BL} Y.Y. Balitzkij, L. N. Lipatov, \textit{Sov. J. Nucl. Phys.}
\textbf{28} (1978) 822; L.N. Lipatov, Zh. Eksp. Teor. Fiz.
\textbf{90} (1986) 1536 [\textit{Sov. Phys.} JETP \textbf{63}
(1986) 904]; E.A. Kuraev, L.N. Lipatov, V.S. Fadin, \textit{ibid}
\textbf{72} (1977) 377 [\textbf{45} (1977) 199].

\bibitem{FJKLPP} R. Fiore, L. Jenkovszky, E. Kuraev, A. Lengyel, F. Paccanoni,
A. Papa, \textit{Phys. Rev.} \textbf{D 63} (2001) 056010.

\bibitem{DATA} Computer readable data files are available
at http://pdg.lbl.gov.

\bibitem{KKL} J. Kontros, K. Kontros, A. Lengyel, Pomeron models and exchange
degeneracy of the Regge trajectories, hep-ph/0006141.

\bibitem{DGMP} P. Desgrolard, M. Giffon, E. Martynov, E. Predazzi, \textit{Eur. Phys. J.}
\textbf{C18} (2001) 555.

\bibitem{BN} B. Nicolescu, \textit{Nucl. Phys. Proc. Suppl.} \textbf{99} (2001) 47.

\end{thebibliography}
\end{document}